\documentstyle[12pt]{article}
\begin{document}

\begin{titlepage}
\thispagestyle{empty}
\begin{flushright} {\small UMDGR--94--089}\\
                   {\small gr--qc/9403017}\\
\end{flushright}

\bigskip

\begin{center}
{\bf \huge Collapse of Kaluza-Klein Bubbles}\\[2ex]

\bigskip

Steven Corley\footnote{corley@umdhep.umd.edu}
and Ted Jacobson\footnote{jacobson@umdhep.umd.edu}\\
\medskip

{\small Department of Physics, University
of Maryland, College Park, MD 20742--4111}\\

\vfill

\begin{abstract}
{\tenrm\baselineskip=12pt
 \noindent

Kaluza-Klein theory admits ``bubble" configurations,
in which the circumference of the fifth dimension shrinks to
zero on some compact surface. A three parameter family of
such bubble initial data at a moment of time-symmetry
(some including a magnetic field)
has been found by Brill and Horowitz,
generalizing the (zero-energy) ``Witten bubble" solution.
Some of these data have negative total energy.
We show here that all the negative energy
bubble solutions start out expanding away from the
moment of time symmetry, while the positive energy
bubbles can start out either expanding or contracting.
Thus it is unlikely that the negative energy bubbles
would collapse and produce a naked singularity.}

\end{abstract}
\end{center}

\vfill

\end{titlepage}
\pagebreak

\section{Introduction}
Negative energy initial data at a moment of time-symmetry
have been found\cite{BrillPfis,BrillHor}
in five-dimensional Kaluza-Klein theory,
with the spatial topology ($S^2\times R^2$)
of the (zero-energy) ``Witten bubble" solution\cite{Witten}.
Since negative mass solutions
are typically associated with naked singularities, one
naturally wonders whether the time evolution of the negative
energy bubble data might produce a naked singularity.
If the bubble expands, this seems unlikely, however if it
were to collapse, the formation of a naked singularity
would seem quite plausible.

Another reason to ask if negative energy bubbles can collapse
is related to the second law of black hole thermodynamics.
If one had a localized negative energy configuration it could be
tossed into a black hole, thus decreasing the mass of the hole.
Although we have not thought carefully about black hole entropy
in Kaluza-Klein theory, it is plausible that decreasing the mass
would decrease the entropy of the hole, thus
violating the classical second law
of black hole mechanics. If the bubble were to expand fast enough
however, it would not be possible to get it into the black hole,
so the existence of this negative energy configuration would
not threaten the second law.

The Witten bubble solution is known in closed form for all time
(it is just an analytic continuation of the five dimensional
Schwarzschild solution), and is known
to expand exponentially away from a moment of time symmetry.
In this paper we will examine the class of initial data found
by Brill and Horowitz in Ref. \cite{BrillHor}, for bubbles in
both vacuum and electrovac Kaluza-Klein theory.
Although the time evolution of these
data is not known
in closed form, we will establish that all the negative energy
bubble solutions
at least {\it start out} expanding, while the positive energy
bubbles can start out either expanding or contracting.

Our conventions are those of Ref. \cite{Wald}.

\section{Bubble data}

All the bubbles we will discuss have three-dimensional spherical
symmetry, in addition to translational symmetry in the fourth,
compact, spatial dimension. Thus, the initial
spatial line element has the form
\begin{equation}
{}^{(4)}ds^2=U\, d\sigma^2 + g_{rr}\, dr^2
+ g_{\theta\theta}\, (d\theta^2 + sin^2\theta d\phi^2),
\label{bubble}
\end{equation}
where $U$, $g_{rr}$ and $g_{\theta\theta}$ are functions of $r$ only.
The circumference of the compact dimension approaches a
constant value  at large $r$.
As we come in from infinity, this circumference changes with $U$,
and eventually $U$
shrinks to zero at $r=r_+$, where the ``bubble" is said to be located.
The manifold is smooth at the bubble provided
the period $P$ of
the $\sigma$ coordinate is chosen so as to avoid a conical singularity
there.
Values of $r$ less than $r_+$ are not relevant---the manifold has a
``hole" in it. For a fixed value of $\sigma$, the 3-topology
is $R^3$ minus an open ball. For fixed $\theta$ and $\phi$,
one has the $r$-$\sigma$ plane, in the shape of a semi-infinite
``cigar". Thus the spatial topology is $S^2\times R^2$.

The bubble solution exhibited by Witten\cite{Witten} is obtained by
analytic continuation of the five dimensional Euclidean
Schwarzschild solution:
\begin{equation}
ds^2=(1-\frac{R^2}{r^2})\, d\sigma^2+(1-\frac{R^2}{r^2})^{-1}\, dr^2
+r^2\, \Bigl(d\chi^2 + sin^2\chi\,
(d\theta^2+sin^2\theta d\phi^2)\Bigr),
\label{es}
\end{equation}
where $\sigma$ is a periodic coordinate of range $2\pi R$.
At large $r$ (\ref{es}) approaches the standard Kaluza-Klein
vacuum $M^4\times S^1$, where $M^4$ is Minkowski spacetime, with
spatial topology $R^3\times S^1$. The $\chi=\pi/2$ slice of
(\ref{es}) is an extremal hypersurface with the bubble topology
$S^2\times R^2$ described above, with the bubble ``surface" located
at $r=R$. Witten observed that the Euclidean Schwarzschild solution
thus describes an instanton mediating decay of the standard
Kaluza-Klein vacuum, via a topology changing tunneling process,
to the zero energy bubble configuration. The classical, Lorentzian
evolution of the bubble solution starting from the initial data
given by the $\chi=\pi/2$ slice of (\ref{es}) is obtained
by the analytic continuation $\chi\rightarrow it + \pi/2$. This yields
the exponentially expanding bubble solution
\begin{equation}
ds^2=-r^2 \, dt^2+(1-\frac{R^2}{r^2})\, d\sigma^2
+(1-\frac{R^2}{r^2})^{-1}\, dr^2
+r^2 cosh^2t\, (d\theta^2+sin^2\theta d\phi^2),
\label{exb}
\end{equation}
which is obviously symmetric in time about $t=0$.

The vacuum bubble data of Brill and Horowitz\cite{BrillHor}
are obtained from the
{\it four}-dimensional Euclidean Reissner-Nordstr\"om solution.
Since the electromagnetic stress-energy tensor is tracefree
in four dimensions, the Ricci scalar vanishes for these metrics,
so they are solutions of the initial value constraint equations
for the {\it five}-dimensional Einstein equation with vanishing
extrinsic curvature. The electrovac data of \cite{BrillHor}
provide a one parameter generalization of the vacuum data, including
a 5-dimensional magnetic field given by the vector potential
$A=k(r_+^{-2}-r^{-2})\, d\sigma$ with $k$ a constant.

These bubble data are of the form
(\ref{bubble}), with $g_{rr}=U^{-1}$, $g_{\theta\theta}=r^2$,  and
\begin{equation}
U=1-\frac{2m}{r}+\frac{b}{r^2}-\frac{4k^2}{3r^4},
\label{U}
\end{equation}
with $m$ and $b$ constants.
The total ADM energy of the corresponding
spacetimes is given by $m/2$\cite{BrillHor,DeserSol} (in units
with the reduced 4-dimensional gravitational coupling constant $^4G$
equal to unity.)

The bubble surface, if it exists, is given by the
largest root of the equation $U(r_+)=0$.
If there is a bubble,
values of $r$ less than $r_+$ are not part of the spacetime.
Regularity of the geometry at $r_+$ requires that the period
$P$ of $\sigma$ be given by $P=4\pi/U'(r_+)$.
In the vacuum case ($k=0$)
one finds $r_+=m+\sqrt{m^2-b}$. No simple expression for
$r_+$ exists in the electrovac case.

For the electrovac case $k^2\ne 0$,
(\ref{U}) yields a bubble for all $m$ and $b$.
For the vacuum case $k^2=0$, it is bubbular for $m>0$
provided $b<m^2$, and for $m\le 0$ provided $b<0$.
The $m=0$, $k^2=0$ data corresponds to the Witten bubble, with $b=-R^2$.
For $m>0$, $U$ increases monotonically from
zero at the bubble to unity at infinity.
For $m<0$,
$U$ increases monotonically from zero at the bubble
to a maximum, then decreases to unity at infinity.
(In the vacuum case, the maximum is equal to $1+ |m^2/b|$,
and is located at $r=|b/m|$.)
For $m=0$, $U$ is monotonically increasing if $b\le 0$, and has
a maximum if $b>0$.

\section{Collapsing or expanding?}

As we do not possess the time dependent solutions with
initial data given by (\ref{bubble}) and (\ref{U}), we shall
presently determine only whether the bubble starts out collapsing
or expanding from the moment of time symmetry. Since it is stationary
at that moment, this amounts to determining
the sign of the second time derivative of the bubble area $A$ evaluated
at the initial moment.

The bubble surface is located where the Killing field generating
rotations around the fifth dimension vanishes. Now in general, a submanifold
defined by the zeros of a Killing field is totally geodesic.
Therefore, by spherical symmetry, the worldline of a point at fixed angles
on the bubble surface must be a geodesic.
For convenience, let us choose Gaussian normal coordinates
$(t,r,\theta,\phi,\sigma)$, with $t=0$ the moment of time symmetry.
Then such a geodesic on the bubble stays at fixed $r$, and
$t$ measures proper time along it.

If we write the angular part of the line element as
$g_{\theta\theta}(r,t)\, (d\theta^2+sin^2\theta d\phi^2)$, the
area of the bubble is given by
$A(t)=4\pi g_{\theta\theta}(r_+,t)$.
(At the moment of time symmetry we have
$g_{\theta\theta}(r,0)=r^2$.)
We thus have
\begin{equation}
\ddot{A}
=4\pi g_{\theta\theta,tt},
\label{ddotA}
\end{equation}
where the partial derivative is to be evaluated at the bubble surface,
and the dot indicates derivative with repect to proper time along the
bubble surface world line.

There are now several ways one can determine $g_{\theta\theta,tt}$
at the moment of time symmetry.
For instance, it is straightforward to simply
write out the Einstein equation in Gaussian normal coordinates.
Equivalently, one may use the standard ADM
evolution equations with unit lapse and vanishing shift.
But the most convenient and informative technique is probably
to employ the geodesic deviation equation, applied to
the radial geodesic congruence at the bubble surface. This tells us that,
since the geodesics remain at fixed $r,\theta,\phi,\sigma$, the
second covariant derivative of their separation distance $L$ is
given by $\ddot L=-R^\theta{}_{t\theta t}\; L$, where, e.g.,
$L=g^{1/2}_{\theta\theta}\, d\theta$ for $d\phi=0$.
We thus obtain
\begin{eqnarray}
g_{\theta\theta,tt}&=&-2 R_{\theta t\theta t}\nonumber \\
&=&2 (R_{\theta\theta}- {}^{(4)}R_{\theta\theta})\nonumber\\
&=&2(T_{\theta\theta}-\frac{1}{3}T g_{\theta\theta}-
{}^{(4)}R_{\theta\theta}),
\label{eom}
\end{eqnarray}
where ${}^{(4)}R_{ij}$ is the Ricci tensor of the spatial metric at
the moment of time symmetry. In the second equality we have used the
fact that on a hypersurface
of vanishing extrinsic curvature, the intrinsic curvature tensor
is equal to the projected spacetime curvature.
In the third equality, we have used
the 5-dimensional Einstein equation
$R_{ab}-\frac{1}{2}R g_{ab}=T_{ab}$.
The problem is now reduced to simply evaluating (\ref{eom})
using the bubble initial data.

For the action
$\int (-g)^{1/2} (R-F^2)$ used by Brill and Horowitz, one finds
$T_{ab}=2F_{ac}F_b{}^c-\frac{1}{2}g_{ab}F^2$.  Since the data
for $F_{ab}$
includes only a magnetic field component $F_{r\sigma}=-2k/r^3$,
one finds at the moment of time symmetry
$R_{\theta\theta}
=-\frac{2}{3}g_{\theta\theta}F^2_{r\sigma}
=-\frac{8}{3}k^2/r^4$.
On the other hand, for the bubble metrics
(\ref{bubble}), (\ref{U}) of interest, we have
${}^{(4)}R_{\theta \theta}|_{t=0}=[r(1-U)]_{,r}=b/r^2-4k^2/r^4$.
Thus (\ref{ddotA}) together with (\ref{eom}) yields
\begin{eqnarray}
\ddot{A}
=&8\pi (-\frac{b}{r_+^2}+\frac{4k^2}{3r_+^4})\nonumber\\
=&8\pi(1-\frac{2m}{r_+}).
\label{final}
\end{eqnarray}

We conclude from (\ref{final}) that the negative energy bubbles all
begin expanding from a moment of time symmetry.
Positive energy bubbles either start out expanding or contracting,
depending on whether $r_+$ is greater than or less than $2m$ respectively.
Unfortunately, our analysis provides no insight into ``why" the negative
energy bubbles start out expanding, and we do not know whether the
result generalizes beyond
the particular initial data we have considered.

Note that we have {\it not} determined whether a bubble that
starts out expanding (or contracting) will continue to do
so for all time. Could a bubble
start out expanding, come to rest, and then turn around
and collapse? Or could a bubble collapse to a minimum size,
and then bounce back? It might seem that our
results already show this is not possible, since the maximum
of expansion or minimum of contraction would be another moment
of time symmetry, which we have already analyzed. However, the data
at the new moment of time symmetry need not be in the three parameter
family found by Brill and Horowitz. For example, in the vacuum case,
all we know is that $R_{ab}=0$, ${}^{(4)}R=0$, and the total energy
is conserved. We do not know however if this fixes the sign of
${}^{(4)}R_{\theta\theta}$ at the bubble.
According to equation (\ref{eom})
this sign is what determines whether
the bubble expands or contracts.

\vskip 1cm

It is a pleasure to thank Dieter Brill for
helpful comments on a draft of this paper.
This work was supported in part by NSF Grant PHY91-12240.

\end{document}